\documentclass[prl,twocolumn,superscriptaddress]{revtex4}
\usepackage[caption=false, justification=centerlast]{subfig}
\usepackage{graphicx}	
\usepackage[latin1]{inputenc} 
\usepackage[T1]{fontenc} 
\usepackage[english]{babel}

\usepackage{lipsum}
\usepackage{bbold}
\usepackage{bm}

\newcommand{\Tr}{\mathrm{Tr}}

\newcommand{\HH}{\mathcal{H}}

\begin{document}

\title{Spin Entanglement and Magnetic Competition via Long-range Interactions in Spinor Quantum Optical Lattices}

\author{Karen Lozano-M\'endez } 
\affiliation{ 
Instituto de F\'\i sica, LSCSC-LANMAC, Universidad Nacional Aut\'onoma de M\'exico, Ciudad de M\'exico 04510, M\'exico}
\author{Alejandro H. C\'asares} 
\affiliation{ 
Instituto de F\'\i sica, LSCSC-LANMAC, Universidad Nacional Aut\'onoma de M\'exico, Ciudad de M\'exico 04510, M\'exico}
 \author{Santiago F. Caballero-Ben\'itez}
 \email{scaballero@fisica.unam.mx}
\affiliation{ 
Instituto de F\'\i sica,  LSCSC-LANMAC, Universidad Nacional Aut\'onoma de M\'exico, Ciudad de M\'exico 04510, M\'exico}

\begin{abstract}
Quantum matter at ultra-low temperatures offers a testbed for analyzing and controlling desired properties in strongly correlated systems. Under typical conditions the nature of the atoms fixes the magnetic character of the system.  Beyond classical light potentials leading to optical lattices and short range interactions, high-Q cavities introduce novel dynamics into the system via the quantumness of light. Here we propose a theoretical model and we analyze it using exact diagonalization and density matrix renormalization group simulations. We explore the effects of cavity mediated long range magnetic interactions and optical lattices in ultracold matter. We find that global interactions modify the underlying magnetic character of the system while introducing competition scenarios. Antiferromagnetic correlated bosonic matter emerges in conditions beyond to what nature typically provides. These allow new alternatives toward the design of robust mechanisms for quantum information purposes, exploiting the properties of magnetic phases of strongly correlated quantum matter.
\end{abstract}

\maketitle

Magnetic quantum matter in optical lattices offers a collection of interesting phenomena in terms of quantum simulation~\cite{Lewenstein}. There are possible applications ranging from quantum computing protocols to quantum system design. These designs could help understand the underlying mechanisms that trigger different kinds of order in analog real materials. The matter is controlled with flexibility, generating effective synthetic quantum matter solids.  The degree of precision achieved allows to control the emergence of different quantum many-body phases. In these settings, strong quantum correlations are present, while paradigmatic escenarios of condensed matter systems regarding quantum phase transitions (QPT) are reproduced. Recent advances controlling ultracold matter allow the experimental realization of fermionic antiferromagnets~\cite{Bakr, Greiner1,Bloch-F,Esslinger-F}. These findings contribute to understand via quantum simulation some features linked to high-Tc superconductivity. In the typical setting, the light fields act parametrically like classical waves generating a ``classical'' optical lattice (COL). The state of the photons is not altered by the back-action of matter. Going beyond classical light fields by the inclusion of cavity back-action in an ultracold system, takes matter into new regimes.  Correlations induced by the high-Q cavity light to the matter and vice-versa modify significantly the energy manifolds experienced by the matter~\cite{HelmutRev2}. Consequently, new correlated phases of matter can emerge. Ultracold systems without COL inside high-Q  cavities with magnetic properties, have been recently achieved by several groups with bosons and fermions~\cite{Esslinger-S, Esslinger-S2, Esslinger-S3,Lev-S, Thompson,Roux,Roux1}. Several proposals regarding exploiting these magnetic interactions without a lattice have been put forward with neutral atoms~\cite{Piazza,Buca,Nunnenkamp,Li}. In this limit, the interplay between cavity light and internal degrees of freedom has been studied in combination with the dynamical and dissipative nature of the system.  Recently, the inclusion of COL and cavity back-action was achieved~\cite{Esslinger-LS,Hemmerich-LS,Esslinger-LS2,Esslinger-LS3}. In these experiments, the competition of different  spatial orders is possible and COL can be controlled arbitrarily. Several studies  have explored  these setups and QPT's~\cite{QOL-Caballero, Nishant,Batrouni, Morigi1, QPT-1, Lode,BDMFT, Helmut,FiniteT}.  However, the interplay regarding magnetism in COL with cavity induced interactions, strong quantum correlations and insulating states, remains largely unexplored.  

\begin{figure}[b!]
\includegraphics[width=0.45\textwidth]{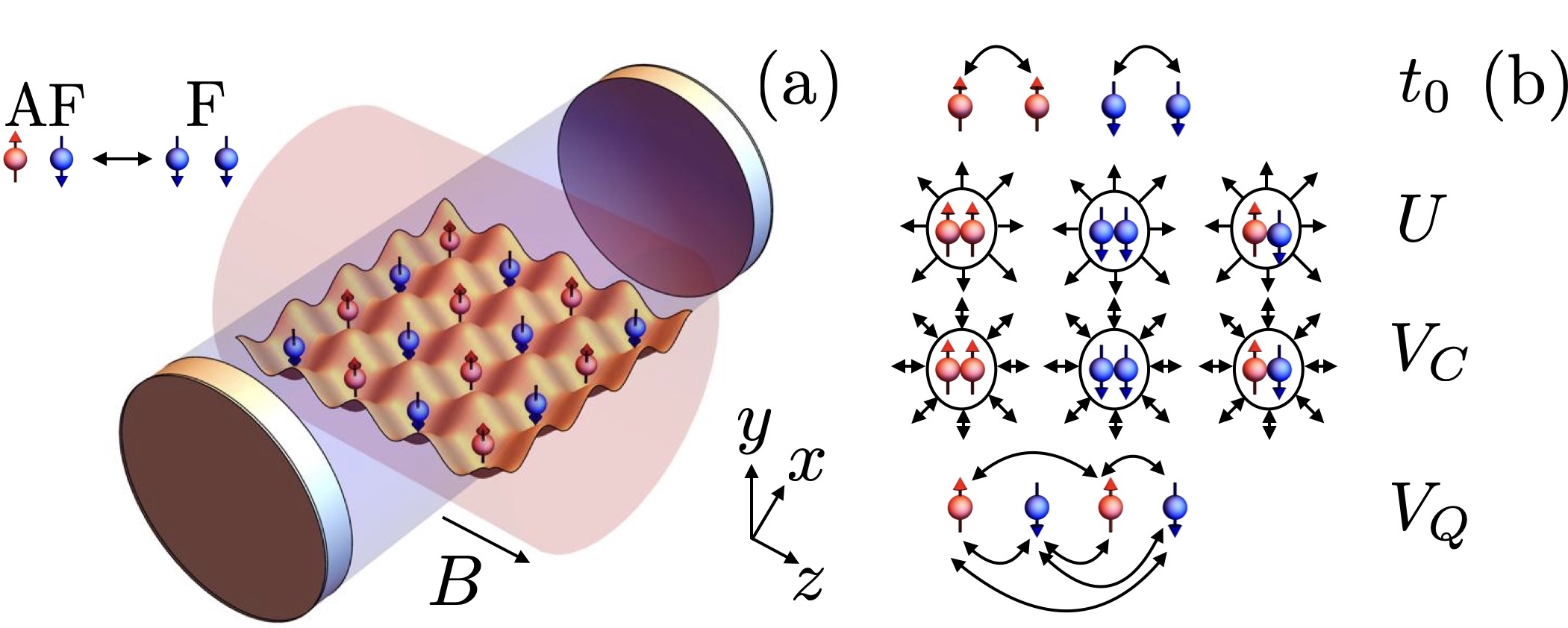} 
\caption{{{ \footnotesize{ {\it Schematic of the system of ultracold atoms in a high-Q cavity with a COL and magnetic degrees of freedom}. (a) Typical atomic Antiferromagnet [AF].  The $V_{\mathrm{OL}}$ (lattice) with intra-cavity light $\hat a$ (cavity axis shade),  pumped light (transverse shade) $\Omega_{z,p}$, and applied magnetic field $B$.  (a, top left)  QPT's between AF$\leftrightarrow$F are possible. (b) Effective atomic interaction processes for different spin componentes: tunneling amplitude $t_0$, on-site repulsion $U$ , intrinsic  (local) magnetic interaction $V_C$,  and cavity induced (global) magnetic interactions $V_Q$.}}}}
\label{fig1}
\end{figure}

Here we show how the interplay of magnetism, COL and cavity induced magnetic interactions allow to control of the emergence of non-trivial magnetic phases of quantum matter. Light and matter are entangled via the cavity generating effective magnetic global interactions. Therefore, quantum antiferromagnets  (AF) and ferromagnets (F) can be produced efficiently. Thus, quantum state engineering of magnetic states with strong correlations can be optimized in a single setup and go beyond what the nature of the atoms typically allows.  

{\it Effective Spinor Quantum Optical Lattice model.} 
We study ultracold bosonic atoms with $F=1$ spin  $\sigma\in\{\downarrow,0,\uparrow\}$,  trapped in a COL subject to a constant magnetic field such that the magnetic sub-levels split inside a high-Q cavity. The alkali atoms in the COL have tunneling amplitude $t_0$, on-site repulsion with strength $U$ and local magnetic interactions (classical) $\propto V_C$. The Hamiltonian describing these processes without the cavity is the Spinor Bose-Hubbard Hamiltonian ~\cite{Ueda,Lewenstein}. We refer to it as ``spinor classical optical lattice'' (SCOL) as the lattice potential comes from a classical treatment of light. The model is: 
\begin{equation}
\HH_{\mathrm{SCOL}}=\HH_U+\frac{V_C}{2}\sum_i\big(\hat{\mathbf{S}}^2_i-2\hat n_i\big),
\end{equation}
 with 
$
\HH_U=-t_0\sum_{\sigma,\langle i, j\rangle}\big(\hat b^\dagger_{i,\sigma}\hat b^{\phantom{\dagger}}_{j,\sigma}+\mathrm{H.c.}\big)+\frac{U}{2}\sum_i\hat n_i(\hat n_i-1)
$, 
The spin operators for $F=1\textrm{ per site are }\hat S_{\nu,i}= \sum_{\xi,\xi'}\hat{b}^\dagger_{\xi,i}F^{\nu}_{\xi,\xi'}\hat{b}^{\phantom{\dagger}}_{\xi',i}, \textrm{ where }\nu\in\{x,y,z\} \textrm{ and }F^\nu$, the angular momentum matrices. The $b_{\sigma,i}^\dagger\;(\hat b_{\sigma,i})$ correspond to bosonic atoms at site $i$ and spin $\sigma$ in the COL. The total spin per site is $\hat{\mathbf{S}}^2_i=\sum_\nu\hat S_{\nu,i}^2$ and the particle number per site operator is $\hat n_i=\sum_\sigma\hat n_{\sigma,i}$.  Additionally, the atoms are inside a single-mode high-Q cavity with the mode frequency $\omega_c$ and decay rate $\kappa$ in off-resonant scattering Fig.\ref{fig1}. Linearly polarized laser light is pumped into the cavity with the Rabi frequency  $\Omega_{z,p}(\mathbf{B})$ dependent on the applied magnetic field $\mathbf{B}=B\hat{e}_z$~\cite{supp} and frequency $\omega_p\;(\Delta_c=\omega_p-\omega_c)$.  The atoms are illuminated from an axis perpendicular to the cavity axis in a standing wave configuration. Each spin component couples with the cavity mode via the effective coupling strength $\tilde{g}_z= g J_z\Omega_{z,p}\sqrt{N_s}/\Delta_a$, with the light-matter coupling coefficient $g$, and the detuning between the light and atomic resonance $\Delta_a=\omega_p-\omega_a$ \cite{Wojciech}. In the COL basis (Wannier basis), the atoms experience the projection of the cavity light-mode with amplitude $J_z\textrm{ over }N_s$ sites~\cite{supp}. For simplicity, the COL is deep enough such that cavity-induced tunneling amplitudes (long range bond processes) are neglected and only COL nearest neighbor tunneling remains~\cite{QOL-Caballero,Bond-Caballero}. Experimentally, this is possible in the non-magnetic version of our system~\cite{Esslinger-LS}.  The Hamiltonian of the light-matter system  is  $\HH=\HH_{\mathrm{SCOL}}+\HH^{a}+\HH^{ab}+\HH_B$. The cavity light Hamiltonian is  $\HH^{a}=-\hbar\Delta_c\hat a^\dagger \hat a$, the operators $\hat a^\dagger$  ($\hat a$) create (annihilate) photons.  The applied magnetic field term is $\HH_B=\mu_B g_S\sum_i\hat{\mathbf{S}}_i\cdot \mathbf{B}$ with $g_S$ the effective Land\'e factor~\cite{FullAto}. The light-atom magnetic interaction ($\HH^{ab}$) is controlled using the vectorial components of the polarizability encoded in $\Omega_{z,p}$~\cite{FullAto,Esslinger-S,Esslinger-S3,supp}. The light-matter interaction is generalized to the lattice case by expanding in the Wannier basis~\cite{EPJD08,QSim-Caballero}, 
\begin{equation}
\HH^{ab}=\frac{\hbar}{\sqrt{N_s}}\sum_i(\tilde{g}_z\varphi_{z,i}\hat a^\dagger+\tilde{g}_z^*\varphi_{z,i}^* \hat a)\hat S_{z,i}.
\end{equation}
The function $\varphi_{z,i}$ encodes the mode structure of the light into the matter~\cite{supp}. This depends on the pump incidence angle with respect to the cavity axis and COL plane~\cite{QOL-Caballero}. Considering the experimental situation described in~\cite{Esslinger-S} without COL, the couplings  of ``$x$'' and ``$y$'' components of the angular momentum are neglected due to energetic considerations, as $|\hbar\Delta_c|\ll \mu_B B$ ~\cite{supp}. Similar decompositions are possible in analog Fermi systems ($S=1/2$)~\cite{FQOL-Caballero,Gabriel,Gabriel-F}. In general, the spatial structure of the light modes gives a natural basis for collective modes~\cite{QOL-Caballero}. For simplicity,  we neglected the non-vectorial (non-magnetic) contributions of the polarizability. Moreover, we take $|\hbar\Delta_a|\gg{\hbar\kappa,|\hbar\Delta_c|}\;\textrm{to avoid heating and }{\hbar\kappa,|\hbar\Delta_c|}\gg t_0$  to avoid non-adiabatic effects in the atomic lattice dynamics~\cite{NJPhys2015}, and  $\kappa\ll|\Delta_c|$. Non-adiabatic effects are minimized in experiments under these assumptions~\cite{Esslinger-S}.  We adiabatically eliminate the cavity light following~\cite{NJPhys2015}. This amounts to effectively integrate out the light. Qualitatively, $\langle\dot{\hat a}\rangle=0$, then it follows: $\langle\hat a\rangle\sim\sum_i\langle\tilde{g}_z\varphi_{z,i}\hat S_{z,i}\rangle$. Beyond this limit, non-adiabatic effects modify slightly the emergence of superfluid (SF) phases. However, insulating states are robust, but effective renormalization of parameters due to cavity noise effects is needed~\cite{Domokos}.  
We find an effective ``spinor quantum optical lattice'' (SQOL),
\begin{equation}
\HH_{\mathrm{SQOL}}=\HH_{\mathrm{SCOL}}+\frac{V_Q}{N_s}\sum_{i,j} f^\varphi_{i,j}\hat S_{z,i}\hat S_{z,j},
\label{effmodel}
\end{equation}
where $\HH_B$ has being effectively decoupled from the low energy atomic dynamics but fixes the quantization axis.  We call the model ``quantum'' as the cavity induced interaction depends on the quantum state of light and the back action of the quantum state of matter~\cite{QOL-Caballero}. Here $f^\varphi_{i,j}=\textrm{Re}(\varphi^*_{z,i}\varphi_{z,j})$, $V_Q=\hbar\Delta_c|\tilde{g}_z|^2/ (\Delta_c^2+\kappa^2)(1+\kappa_{\mathrm{nad}})$ and non-adiabatic corrections: $\kappa_{\mathrm{nad}}=-\kappa^2/\Delta_c^2$~\cite{Adiabatic}.  $\kappa_{\mathrm{nad}}$ shows that photon loses need to be minimized to control $\mathrm{sign}(\Delta_c)$. Stability of photon steady states ($\Delta_c>0$) is sensitive to atomic deconfinement temperature effects. Metastable states are posible for $\Delta_a>0$. Photon steady states with either sign of  $\Delta_c$ were achieved in ms~\cite{Esslinger-LS3}.     
The matter will self-organize in such a way that the cavity induced interaction components ``$i,j$'' are maximized (minimized) by  $f^\varphi_{i,j}\Delta_c>0\;(f^\varphi_{i,j}\Delta_c<0)$, as cavity light maximizes (minimizes) akin to superradiance (subradiance).  For minimized cavity light, quantum fluctuations play a fundamental role~\cite{QOL-Caballero,QSim-Caballero}. The number of photons in the cavity is 
$n_{\mathrm{ph}}=\langle \hat a^\dagger\hat a\rangle\approx|V_Q/(N_s\Delta_c)\sum_{i,j}f^\varphi_{i,j}\langle \hat S_{z,i}\hat S_{z,j}\rangle|$. 
In addition, there is competition between the typical local (short-range) processes in the Bose-Hubbard model ($\propto U\;\textrm{and}\;t_0$), ``local'' spin classical interactions ($\propto V_C$) and the ``global'' (long-range) cavity induced spin quantum interactions ($\propto V_Q$). Their interplay leads to different quantum critical points (QCP's). Typical frequencies of the analogous system without COL are $\kappa\ll|\Delta_c|\sim \mathrm{MHz}\; \textrm{and}\; E_R\sim h\times 4\mathrm{kHz}$, $E_R$ the recoil energy~\cite{Esslinger-S}. Typical values of the non-magnetic system with COL are $t_0\sim E_R\;\textrm{and}\;\Delta_a \sim10-100 \textrm{GHz}$~\cite{Esslinger-LS3}.  In the case of SCOL ($V_Q=0$), the sign of the magnetic interaction $V_C$ is fixed by the nature of the atom to be either Ferromagnetic (F) for $V_C<0$ or Antiferromagnetic (AF) for $V_C>0$. Typical atoms used are:$\phantom{0}^{87}\textrm{Rb (F)}$, $\phantom{0}^7\textrm{Li (F)}$ or $\phantom{0}^{23}\textrm{Na (AF)}$ with $V_C/U\sim(-0.005,-0.23,0.04)$~\cite{Ueda}. In the system with $V_Q\neq 0$, parameters can be tuned externally (i.e.  $\Delta_c$) triggering different magnetic behaviour. We study configurations ($\phi_\pm=\varphi_{z,i}=(\pm1)^i$) where the pump incidence angle maximizes diffraction generating homogenous coupling ($\phi_+$) or staggered density coupling in the diffraction minima ($\phi_-$) similar to current experimental settings. Tuning non-insulating  antiferromagnetic (AF) and ferromagnetic (F) states with $\phi_-$ is experimentally  feasible~\cite{Esslinger-S}. More elaborate scenarios and flexibility can be achieved depending on the pumps, the cavity setup and the magnetic field~\cite{QOL-Caballero,QSim-Caballero,HelmutRev2,couplings}. 

{\it Magnetic interactions.}
It is  experimentally possible to prepare the system with different spin populations without the lattice in the cavity~\cite{Esslinger-S}. Without the cavity with COL, the phase diagram is well known~\cite{Lewenstein,Ueda,Rousseau}, having Polar, F and AF phases. We choose commensurate fillings in the lattice to study the behaviour between Mott-insulator (MI) phases driven by $U$ and the magnetic dynamics. In the effective model for convenience,  we introduce linear $\epsilon_{\uparrow\downarrow}\sum_i\hat S_{z,i}$ (quadratic, $\epsilon_0\sum_i\hat n_{0,i}$) magnetic field shift favouring (suppressing) one of the spin components ``$\uparrow\downarrow$'' (``0'') relevant for F (AF) ordering, $\epsilon_\sigma$ a small perturbation~\cite{Ueda}.  
In the case where $V_C=0$ the behaviour is simplified as the ``0'' component decouples due to the interaction form. The many-body quantum state is $|\Psi\rangle=|\Psi_0\rangle\otimes|\Psi_{\uparrow,\downarrow}\rangle$ ~\cite{supp}. For simplicity in what follows, we consider $\epsilon_0>0$, suppressing Polar configurations. Preparing the system  with the ``$0$'' component empty is experimentally achievable \cite{Esslinger-S}.

{\it Cavity induced ferromagnetic configurations ($V_C=0$).}
 The behaviour is intuitive for $\phi_+$ and $V_Q<0$, the system maximizes either spin component ``$\uparrow\downarrow$'' depending on the sign of $\epsilon_{\uparrow\downarrow}\neq0$ having a ferromagnet.  Similarly,  for $\phi_-$, $V_Q>0$, the system is always F. In these cases, the system behaves as a single component Bose-Hubbard model either ``${\uparrow\downarrow}$''. The ground state is magnetically trivial being fully polarised~\cite{Ferro}. The system goes from F insulator (FI) to F superfluid (FSF) increasing $t_0/U$.

\begin{figure}[t!]
\includegraphics[width=0.48\textwidth]{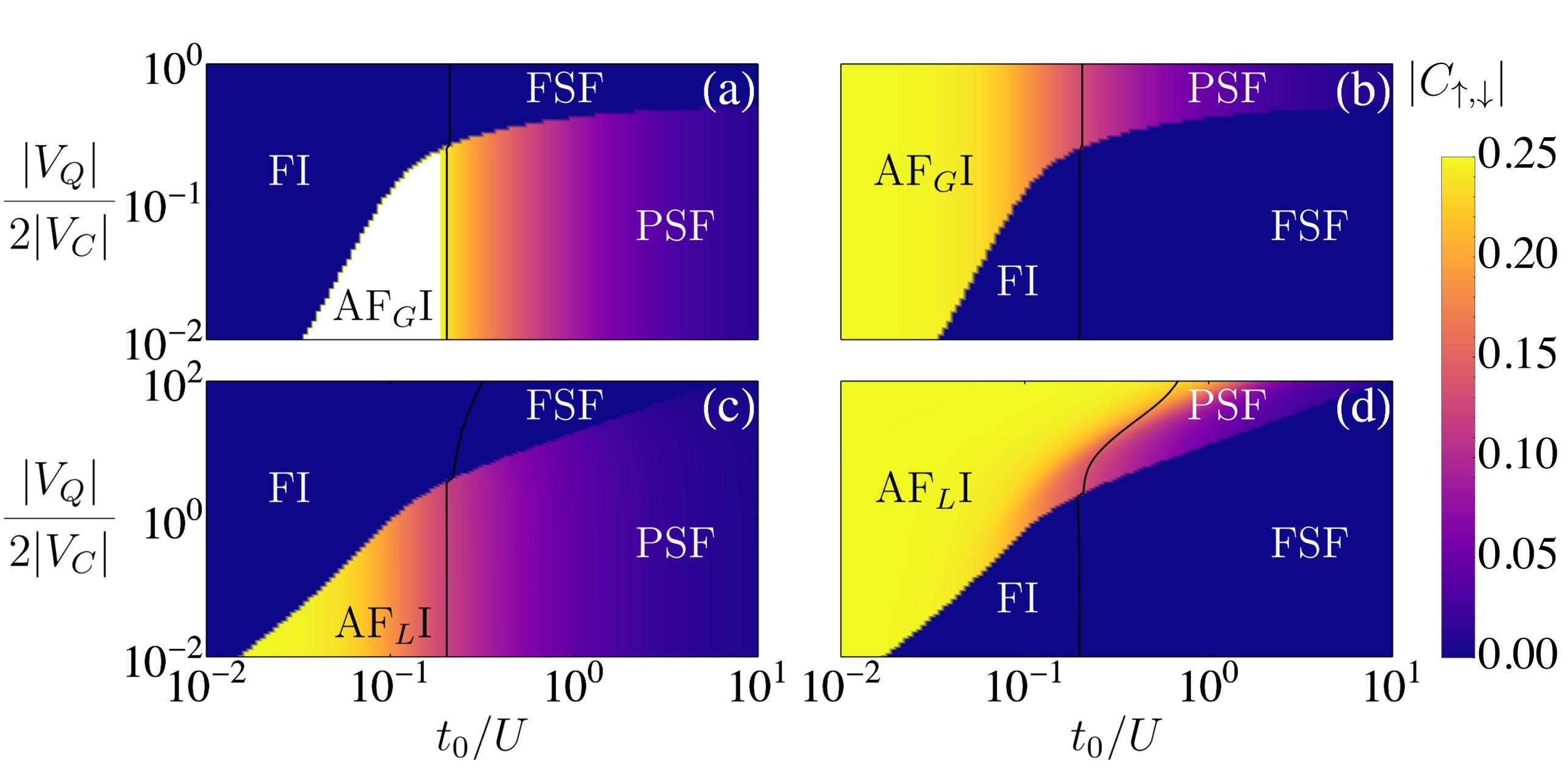} 
\captionsetup{width=0.4\textwidth}
\caption{{\footnotesize{ {\it Phase diagrams of magnetic configurations competition scenarios.} 
The spin quantum correlations $|C_{\uparrow,\downarrow}|$ for $V_Q$(global) and $V_C$(local) magnetic interactions. The competition triggers QPT's between $\mathrm{AF}\leftrightarrow\mathrm{F}$. Black lines approximate the QCP  of  the SF-MI  QPT, where the total on-site number fluctuations are half of the limit $t_0\gg U, \max(\Delta(\hat n_i)^2)=1-N_s^{-1}$. Parameters are: $\textrm{(a) }V_C>0,\;V_Q<0,\;\phi_+; \textrm{(b) } V_C<0,\;V_Q>0,\;\phi_+;  \textrm{(c) } V_C>0,\;V_Q>0,\;\phi_-;\;\textrm{(d) }V_C<0,\;V_Q<0,\;\phi_-, \textrm{with } N_s=8,\; \textrm{2 spin components, }\epsilon_{\uparrow\downarrow}=10^{-8}U$ using ED.}}}
\label{QPD}
\end{figure}

{\it Cavity induced antiferromagnetic correlations ($V_C=0$).}
Notably, if $V_Q<0, \phi_-$  or $V_Q>0, \phi_+$, the situation is not magnetically trivial as AF correlations emerge. The system is a balanced mixture $\sum_{i}\langle\hat n_{\uparrow,i}\rangle=\sum_i\langle\hat n_{\downarrow,i}\rangle$. However, the total population fluctuations per site ($\Delta(\hat n_i)^2=\langle\hat n_i^2\rangle-\langle\hat n_i\rangle^2$) for large $U$ are minimized  as the Mott gap ($\Delta_e=U$) opens. The MI state exhibits large fluctuations per site in the ``${\uparrow\downarrow}$'' components. The system goes from a AF insulator (AFI) to a paramagnetic SF (PSF) as $t_0/U$ increases~\cite{supp}.

{\it Deep Mott Insulator limit AF's ($U\gg t_0$, $V_C=0$).}
We study the spin quantum correlations $C_{\uparrow,\downarrow}=\mathrm{cov}(\hat n_{\uparrow,i},\hat n_{\downarrow,i}),\textrm{ the staggered magnetization }m_\pi$,  the magnetization $m_0;\textrm{ with }\mathrm{cov}(\hat X,\hat Y)=\langle \hat X \hat Y\rangle-\langle \hat X\rangle\langle\hat Y\rangle, m_\theta=\sqrt{\langle| \sum_{i} e^{i\theta d_i} \hat S_{z,i}|^2\rangle}/N_s,\; d_i=i_x+i_y$ and the lattice position $\{i_x,i_y\},\;i_{x/y}\in\mathbb{Z}$. The relation  between fluctuations in this limit with $V_Q>0, \phi_+$ or $V_Q<0, \phi_-$ is:  $\Delta(\hat n_i)^2=0,\; \Delta(\hat n_{{\uparrow\downarrow},i})^2=1/4\textrm{ and }\mathrm{C}_{\uparrow,\downarrow}=-1/4$. In the case of $V_Q>0,\textrm{ and }\phi_+$, the ground state is a degenerate insulator with ``global'' AF correlations ($\mathrm{AF}_G$I) and maximal $|C_{\uparrow,\downarrow}|\neq0$. The  ground state degeneracy is $g^0_G\sim2^{N_s-1/3}(N_s)^{-1/2}$, all the states with magnetization $m_0=0$ and one particle per site.  This large degeneracy persists for small $t_0$~\cite{supp}. The excitation gap is $\Delta_e=\min(U,4V_Q/N_s)$, $m_\pi\to O(N_s^{-1})\textrm{ and }n_{\mathrm{ph}}\to O(N_s^{-1})$ for $N_s\gg1$.  Surprisingly, for $V_Q<0\textrm{ with }\phi_-$,  we find that a ``local'' insulating AF state with degeneracy $g^0_L=2,\;\Delta_e=U,\textrm{ maximal }m_\pi=1\textrm{ and }n_{\mathrm{ph}}\propto N_s^2$, a staggered quantum antiferromagnet, with $m_\pi$ the typical AF order parameter.  The ``local'' AF insulating states ($\mathrm{AF}_L$I) present true conventional AF order and non-trivial magnetic quantum correlations. In contrast, $\mathrm{AF}_G$I has only non-trivial magnetic quantum correlations. Either ground state has ``${\uparrow\downarrow}$'' components anticorrelated \cite{supp}. Thus,  the many-body insulating states with $C_{\uparrow,\downarrow}\neq 0$ are $|\Psi_{\uparrow,\downarrow}\rangle\neq|\Psi_\uparrow\rangle\otimes|\Psi_\downarrow\rangle$ with the entanglement entropy between spin sectors $\mathcal{S}_\sigma\neq0$, as we confirm below. Deep in the MI, these facts are independent of dimensionality.  Away from the MI,  the SF state emerges decreasing spin correlations while reaching a paramagnetic state, as $U\to0$ then $\mathrm{C}_{\uparrow,\downarrow}\to0$ and $|\Psi_{\uparrow,\downarrow}\rangle\approx|\Psi_\uparrow\rangle\otimes|\Psi_\downarrow\rangle$.

{\it Competition of Magnetic Configurations.}
 Interestingly, even if $V_C$ is fixed by nature for a given alkali atom, modifying the pump angle and $V_Q$ allows the competition  between AF and F in a single setup.  The emergent phases of quantum matter can be understood by analysing the number fluctuations  for  ``$\uparrow\downarrow$'' components, the total number fluctuations,  $C_{\uparrow,\downarrow}$ and  $\mathcal{S}_\sigma$. The information of fluctuations and correlations might be accessed via in situ measurements~\cite{Greiner1} or direct measurements of AF correlations~\cite{Hilker,Hart}. 
We perform simulations with exact diagonalization  (ED) and density matrix renormalization group (DMRG) in 1D~\cite{ed,dmrg,supp}. We construct the ground state phase diagrams in Fig. \ref{QPD} (a-d)  with ED (8 sites, 2 spin components, and $\sim5\times10^5$ states).  The F$\leftrightarrow$AF competition for $\phi_{+/-}$ occurs by choosing different/equal signs in  $V_{Q/C}$. The sharp boundaries between F-AF (AF-F) occur being 1st order QPT's, as Hilbert spaces are orthogonal, see below.  The ratio $V_Q/V_C$ determines the passage starting from insulating regions AFI (FI) to have a 1st order transition to a FI (AFI) state in the limit $U\gg t_0$, while FSF (PSF) emerges for $U\ll t_0$.  As a function of the lattice depth (effectively $t_0/U$) at fixed ratios $V_Q/V_C$, the following scenarios are  possible for $\phi_{\pm}$,
$$ 
\mathrm{FI}\leftrightarrow\mathrm{AFI}\leftrightarrow\mathrm{PSF}
\quad\mathrm{or}\quad 
\mathrm{AFI}\leftrightarrow\mathrm{FI}\leftrightarrow\mathrm{FSF}
$$
Using DMRG with up to $\sim100$ sites with 2 spin components and finite size scaling~\cite{supp,dmrg}, we confirm a finite $\Delta_e$ that vanishes at the transition between AF$\leftrightarrow$F phases in general, Fig. 3 (a b). In F and $\mathrm{AF}_L$I  phases $\Delta_e\sim U$. However, from the deep MI limit of $\mathrm{AF}_G$I,  $\Delta_e\sim\min(N_s^{-1},U)$ can be considerably smaller. In the the large $V_Q/U$ limit,  $\Delta_e\sim U$. The AF order parameter, $m_\pi$ with $\phi_-$ decreases as $\Delta_e$ closes, for details~\cite{supp}. In general, via the cavity induced magnetic interactions it is possible to control whichever scenario one would desire. 

\begin{figure}[t!]
\includegraphics[width=0.47\textwidth]{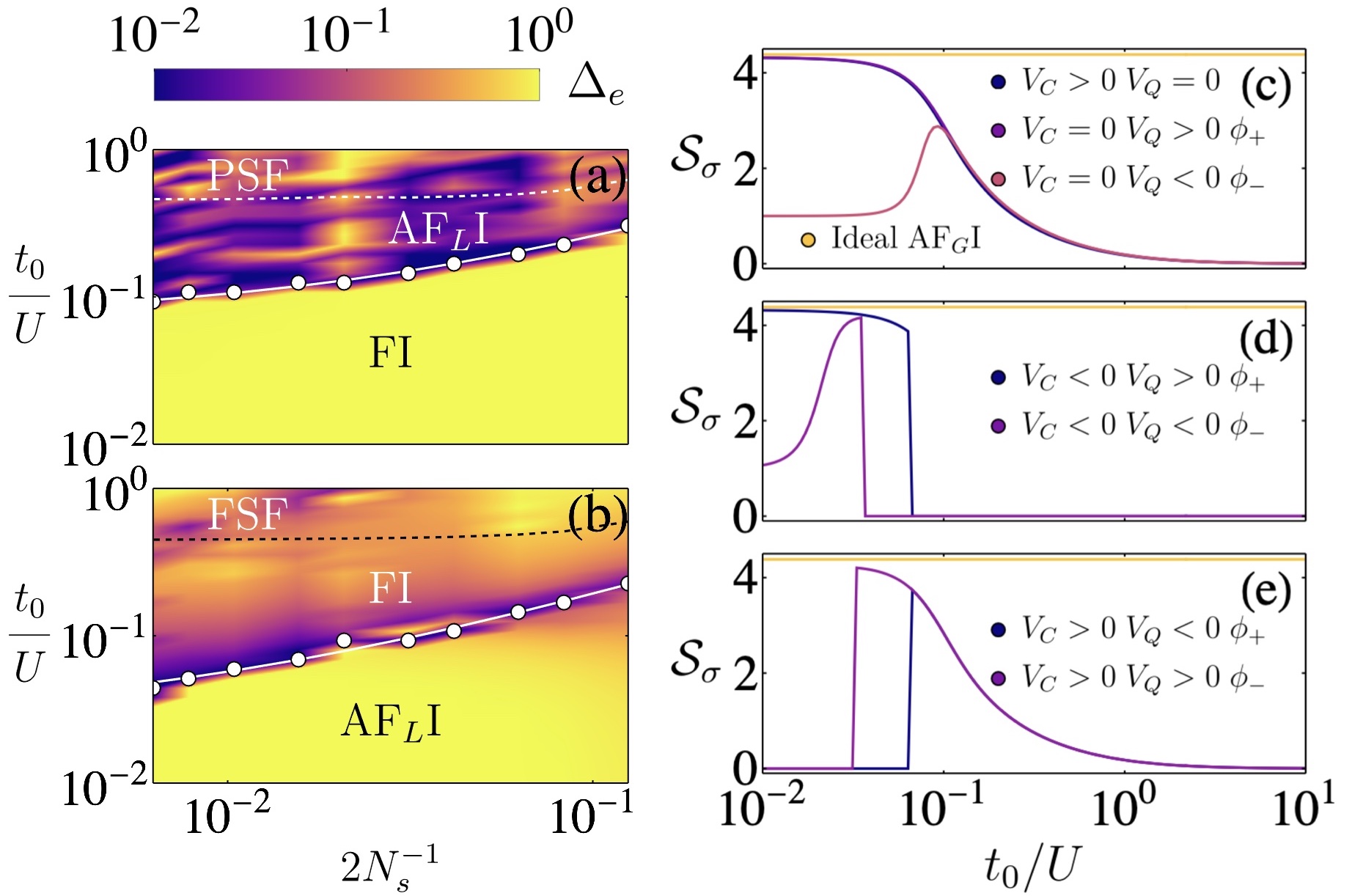} 
\captionsetup{width=0.47\textwidth}
\caption{
{\footnotesize{\it Excitation gap, staggered magnetization, and entanglement entropy of spin components}.
Panels $\textrm{(a,b) }\Delta_e$ from DMRG simulations in 1D with $\phi_-$. White lines are the finite size scaling fits (FFS) for the AF$\leftrightarrow$F  QPT (white points).  FSS for the critical $t_0$,  $t_c\approx \big(a_0+a_1N_s^{-3/4}\big)U$.  Estimated SF-MI boundary, white (black) dashed lines (Similar procedure as in Fig. 1). Parameters are: $\textrm{(a) }V_Q=V_C=0.035U,\;\{a_0,a_1\}=\{0.072,1.567\};\textrm{ (b) }V_Q=V_C=-0.03U,\;\{0.028,1.768\}$, with maximum number of atoms per site $n_{max}=4$, commensurate filling $N_b=N_s$, $\epsilon_{\uparrow\downarrow}=10^{-4}U$.
Panels (c-e), the entanglement entropy $\mathcal{S}_\sigma$. Parameters are: $V_Q\neq0:\;|V_Q|/|V_C|=0.05;\; V_C=0.03 U$, $N_s=6$, $\epsilon_{\uparrow\downarrow}=10^{-8}U$ using ED.}}
\label{Ent}
\end{figure}

{\it Spin Entanglement.}
Typically entanglement partitioning considers spatial subsystems. However, we are interested in how the entanglement between spin projections relates to the magnetic properties of the many-body state. Therefore, we analyze by tracing over different spin projection subsystems via the entanglement entropy $\mathcal{S}_\sigma=-\Tr[\rho_\sigma\log_2\rho_\sigma]$~\cite{RDM}.
We find that $\mathcal{S}_\sigma$ gets maximized in the insulator region of the phase diagram for SCOL:  $V_C>0,V_Q=0$, and SQOL: $V_C=0,V_Q>0,\phi_+$ having $\max(\mathcal{S}_\sigma)=\log_2(g^0_G)$. This is the entanglement entropy of the ideal $\mathrm{AF}_G$I, 
deep in the MI, with maximal $|\mathrm{C}_{\uparrow,\downarrow}|$. The transition is smooth due to dimensionality and finite-size. Surprisingly, this is not the case for  $\mathrm{AF}_L$I with $V_C=0,V_Q<0,\phi_-$. The difference origins in the degeneracy of the ground states deep in the MI. Here $g_L=g_L^0=2$ having $\mathcal{S}_\sigma=1$.  Increasing $t_0/U$, non-monotonic character emerges because the ground state degeneracy increases reaching the MI-SF  transition as $\Delta_e\to0$, while $\mathcal{S}_\sigma$ maximizes. Beyond the QCP, $\mathcal{S}_\sigma$ vanishes as PSF is separable, $|\Psi_{\downarrow,\uparrow}\rangle=|\Psi_\downarrow\rangle\otimes|\Psi_\uparrow\rangle$, with $\langle\hat n_{\uparrow,i}\rangle=\langle\hat n_{\downarrow,i}\rangle\neq0\;\forall i$, Fig.\ref{Ent} (c).

Stands out that for $V_{C/Q}\neq0$ with the competition between magnetic configurations, $\mathcal{S}_\sigma$  shows the 1st order character of the AF$\leftrightarrow$F  QPT, Fig.\ref{Ent} (d,e).  These confirm that AF and F  belong to orthogonal Hilbert space sectors.  Beyond the QPT, in the F side, $\mathcal{S}_\sigma=0$ and $C_{\uparrow,\downarrow}=0$, a completely polarized system
with $\langle\hat n_{\sigma,i}\rangle\neq0,\langle\hat n_{-\sigma,i}\rangle=0\;\forall i$. For AF, spin entanglement and $C_{\uparrow,\downarrow}$ maximize.  
Moreover, QPT occurs for smaller  $t_0/U$ for $\phi_-$ than $\phi_+$.  Via $\mathcal{S}_\sigma$, it is possible to discriminate $\mathrm{AF}_{L/G}$ only for $V_C<0$, Fig. \ref{Ent}(d).

Essentially,  $\mathcal{S}_\sigma\sim4\log_2(f(g_{L/G}))|C_{\uparrow,\downarrow}|$ for some function $f(g)$. Away from the SF-MI QCP, $\mathcal{S}_\sigma\sim4\log_2(g^0_{L/G})|C_{\uparrow,\downarrow}|$.
The behaviour of $\mathcal{S}_\sigma$  clarifies the impact of competition, degeneracy and magnetic correlations in the ground state. 

We conclude that $\mathrm{AF}_G$I's  have more resilient  entanglement  accessible for sufficiently large $V_Q$ at lower lattice depths. This could be useful as a resource for quantum state preparation (Cluster states) in quantum information  (QI) schemes~\cite{CSOL,CSOL2}. This robustness could be exploited in analog experiments to \cite{gates}. Here qubit gates with spinless neutral atoms ($10^{4}$) using the Bose-Hubbard Hamilonian were explored.

Spinor quantum optical lattices offer great flexibility to explore the nature of different magnetic quantum phases of matter. We show that in the simplest setup, the emergence  and competition of correlated antiferromagnetic  or ferromagnetic quantum phases of matter can be investigated.  Moreover, the system naturally supports additional competing orders via the light-induced non-magnetic interaction terms (density wave, multimode, bond)~\cite{QSim-Caballero,Bond-Caballero}. Changing atomic species (i.e. rare-earth atoms~\cite{Ferlaino, RareEarth1}) allows other finite range interactions, density dependent tunneling processes and peer into the landscape of Kondo physics. Using geometrically frustrated AF's~\cite{Frustrated1,Frustrated2}  will generate emergent degrees of freedom and possibly long-range quantum spin liquids~\cite{QSL1, QSL2,QSL3,QSL4}. It should be feasible to explore the interplay with static gauge fields~\cite{GaugeStatic1,GaugeStatic2} and cavity generated spin-orbit coupling via Raman transitions~\cite{GaugeDynamical,Ritsch-SO}. Moreover, a plethora of possibilities using dynamical gauge fields can be considered~\cite{GaugeMore1, GaugeMore2}, exploring high energy physics analogs beyond local field theories.

From the  QI perspective, entanglement can be tailored on demand and it is robust between spin components globally. These suggest new means to manipulate and encode information in the emergent magnetic structures found. QI and topological order~\cite{TopoApplied,TopoApplied2} could be explored further. The combination with measurement allows dynamical order control with passive measurement setups~\cite{Gabriel,Gabriel-F}, the inclusion of feedback protocols to tailor criticality~\cite{Feedback-PT1, Feedback-PT2,Feedback-PT3,Esslinger-FB}, engineering system dynamics ~\cite{Feedback-Gabriel,tbs,Ashida} and to study the interplay with time crystals~\cite{Feedback-PT1,Demler-TC,Cosme, Hemmerich-TC1,Hemmerich-TC2}. 

{\it Note added.-} Recently, we became aware of a paper \cite{dipoles} related to our work in dipole systems with lattices without insulators.

\begin{acknowledgments}
We thank R. J\'auregui, D. Sahag\'un, I. B. Mekhov, A. Chiocchetta, F. Piazza, and T. Donner for useful discussions. This work was supported by the grants UNAM, DGAPA-PAPIIT:  IN109619, UNAM-AG810720, LANMAC-2019 and CONACYT Ciencia B\'asica: A1-S-30934. K. L.-M. thanks DGAPA-UNAM PAPIIT program for financial support. A. H. C. acknowledges financial support from CONACYT. We acknowledge infrastructure support for the computations from the ``Laboratorio de Simulaciones Computacionales para Sistemas Cu\'anticos'' in LANMAC (LSCSC-LANMAC) at IF-UNAM. 
 \end{acknowledgments}

\newpage

\section*{Supplemental Material: Spin Entanglement and Magnetic Competition via Long-range Interactions in Spinor Quantum Optical Lattices}

\textit{The Bose-Hubbard parameters and overlap integrals.} 
The tunneling amplitude of the bosons is $t_0$, the on-site interaction is $U$. The effective parameters linked to the cavity can be calculated using Wannier functions,
\begin{eqnarray}
t_0&=&\int w(\mathbf{x}-\mathbf{x}_i)\left(\frac{\hbar^2}{2m}\nabla^2-V_{\mathrm{OL}}(\mathbf{x})\right)w(\mathbf{x}-\mathbf{x}_j)\mathrm{d}^{n_d} x,
\nonumber\\
\end{eqnarray}
 $w(\mathbf{x})$ are the Wannier functions with $i$, $j$ nearest neighbours. The classical optical lattice potential is $V_{\mathrm{OL}}(\mathbf{x})=V_0 \sum_{k=1}^{n_d}\sin^2(2\pi x_k/\lambda)$ with $n_d$ the dimension. Typically, for a deep lattice $V_0\gtrsim10E_R$, then $t_0\sim 0.1  E_R$,
where $E_R$ is the recoil energy. The light-matter coupling coefficients in the direction $\nu\in{x,y,z}$ are,
\begin{eqnarray}
J_\nu\varphi_{\nu,i}=\int |w(\mathbf{x}-\mathbf{x}_i)|^2u^*_{V,\nu,c}(\mathbf{x})u_{V,\nu,p}(\mathbf{x})\mathrm{d}^{n_d} x,
\end{eqnarray}
where $u_{V,\nu,c/p}$ are the effective cavity/pump mode functions from the vectorial contributions of the polarisation (see supplemental of \cite{Esslinger-S}). For the cases considered here, the effective mode functions are such that the projections in the ``$x$'' and ``$y$'' direction are neglected due to energetics, see below. 

\textit{Approximation of  relevant spin component.}
In principle the components $J_\nu$ that quantify the response with respect to the magnetic field applied depend on it implicitly. One has that the relevant vectorial components of the polarization that affect the single particle atomic dynamics (quadratic Stark shift) to second order perturbation theory are~\cite{Esslinger-S},
\begin{equation}
\Delta H_{V}\propto i\frac{\alpha_V}{2F}\text{\boldmath$\epsilon$}_p\times\text{\boldmath$\epsilon$}_c\cdot\hat{\mathbf{F}}(\hat a^\dagger-\hat a)
\end{equation}
with $\text{\boldmath$\epsilon$}_{p,c}$ the pump/cavity polarizabilities assuming linear polarization and $\hat{\mathbf{F}}$ the angular momentum operators for the ground state. We assume that the magnetic field applied $\mathbf{B}=B\hat{e}_z$, is such that $\mu_B B\gg|\hbar\Delta_c|$ as in the case of the experiment~\cite{Esslinger-S}. As such, the components $F^x$ and $F^y$ couple states with very high energy which leads to a small transition probability in this limit.  Therefore, the populations of states that couple $F^{x/y}$ are negligible. However the $F^z$ component is diagonal~\cite{FullAto}. Setting: $\text{\boldmath$\epsilon$}_p\times\text{\boldmath$\epsilon$}_c\cdot\hat{\mathbf{F}}=\epsilon_p\epsilon_c\hat{F}_z$, with $|\text{\boldmath$\epsilon$}_{p,c}|=\epsilon_{p,c}$. It follows that the single atom contribution to the energy is,
\begin{equation}
\Delta H_{V}\propto( c \hat a^\dagger+c^* \hat a)\hat F_z.
\end{equation}
with $c=i\frac{\alpha_V\epsilon_p\epsilon_c}{2F}$, with $F$ the magnitude of the atomic spin and $\alpha_V$ the reduced dynamical vector scalar polarizability of the atom in the fine-structure level~\cite{FullAto}. Therefore, the effective Rabi frequency $\Omega_{z,p}$ depends implicitly on $\mathbf{B}$.
It is possible with our framework to consider beyond this limit.  In view of simplicity, we left this possibility for future work.

\textit{Many-body Quantum States}. 
The general many-body quantum state of the matter can be written as:
$|\Psi\rangle=\sum_{\bar{\nu}}\alpha_{\bar{\nu}}|\Psi_\downarrow\rangle_{\nu_\downarrow}\otimes|\Psi_0\rangle_{\nu_0}\otimes|\Psi_\uparrow\rangle_{\nu_\uparrow}$ with $|\Psi_\sigma\rangle_{\nu_\sigma}=|n_{\sigma,1}\rangle\otimes\cdots\otimes|n_{\sigma,N_s}\rangle_{\nu_\sigma}$, where $\bar{\nu}=\{\nu_\downarrow,\nu_0,\nu_\uparrow\}$ denotes an element of the Hilbert space  basis with amplitude $\alpha_{\bar{\nu}}$. A particular combination of quantum numbers $\{n_{\sigma,1},\dots,n_{\sigma,N_s}\}$ for each spin component $\sigma$ is denoted by $\nu_\sigma$. In the case where the ``0'' of the components is separable, we have: $\alpha_{\bar{\nu}}=\alpha_{\nu_0}\alpha_{\nu_\downarrow,\nu_\uparrow}=\alpha_{\nu_0} \alpha_{\tilde{{\nu}}}$ with $\tilde{{\nu}}=\{\nu_\downarrow,\nu_\uparrow\}$. The state is: $|\Psi\rangle=|\Psi_0\rangle\otimes|\Psi_{\downarrow,\uparrow}\rangle$, with $|\Psi_0\rangle=\sum_{\nu_0}\alpha_{\nu_0}|\Psi_0\rangle_{\nu_0}$ and $|\Psi_{\downarrow,\uparrow}\rangle=\sum_{\tilde{{\nu}}}\alpha_{\tilde{{\nu}}}|\Psi_\downarrow\rangle_{\nu_\downarrow}\otimes|\Psi_\uparrow\rangle_{\nu_\uparrow}$. A completely separable state in the magnetic component sense (deep in the SF) is: $|\Psi\rangle=|\Psi_0\rangle\otimes|\Psi_\downarrow\rangle\otimes|\Psi_\uparrow\rangle$, with $|\Psi_\sigma\rangle=\sum_{\nu_\sigma}\alpha_{\nu_\sigma}|\Psi_\sigma\rangle_{\nu_\sigma}$, then spin components are not correlated between each other.

Deep in the MI regime ($t_0\ll U$) for $V_C=0$ with the ``0'' component empty or decoupled, we have the following structure for ground states |$\Psi_{\downarrow,\uparrow}\rangle$ : the ``global'' AF ($\mathrm{AF}_G\mathrm{I}$) is 
$$
\begin{array}{l}
\cdots\;
|\cdots\uparrow\uparrow\downarrow\downarrow\cdots\rangle\;
|\cdots\uparrow\downarrow\uparrow\downarrow\cdots\rangle\;
|\cdots\uparrow\downarrow\downarrow\uparrow\cdots\rangle\;
\phantom{\cdots}
\\
\phantom{\cdots}\;\,
|\cdots\downarrow\uparrow\uparrow\downarrow\cdots\rangle\;
|\cdots\downarrow\uparrow\downarrow\uparrow\cdots\rangle\;
|\cdots\downarrow\downarrow\uparrow\uparrow\cdots\rangle\;
\cdots,
\end{array}
$$
the $g_G\sim2^{N_s-1/3}(N_s)^{-1/2}$ possible combinations, $N_s$ the number of sites; similarly for the ``local'' AF ($\mathrm{AF}_L\mathrm{I}$):
$$
\; |\cdots\uparrow\downarrow\uparrow\downarrow\cdots\rangle\;\\\; \
\;\;
|\cdots\downarrow\uparrow\downarrow\uparrow\cdots\rangle\;,
$$
with $|\uparrow\rangle=|0,1\rangle$ and $|\downarrow\rangle=|1,0\rangle$ in the Fock space representation $|n_{i,\downarrow},n_{i,\uparrow}\rangle$. It is easy to see that $\langle \hat n_{\uparrow,i}\hat n_{\downarrow,i}\rangle=0$ and $\langle \hat n_{\sigma,i}\rangle=1/2$, leading to $C_{\uparrow,\downarrow}=-1/4$.

\begin{figure}[t!]
\includegraphics[width=0.41\textwidth]{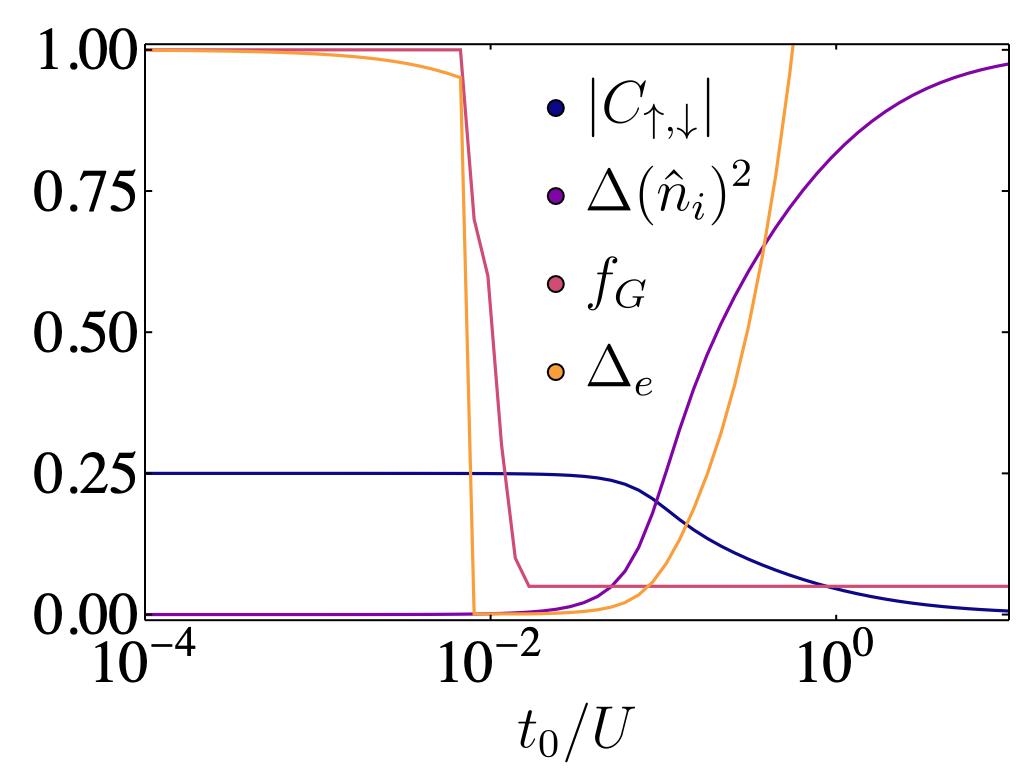} 
\captionsetup{width=0.41\textwidth}
\caption{
{\footnotesize{ 
{\it Exact Diagonalization simulations for $V_C=0$, $V_Q>0$ and $\phi_+$.}
Total number fluctuations $\Delta(\hat{n})^2$, the quantum covariance $|C_{\uparrow,\downarrow}|$, the scaled degeneracy of the ground state $f_G=g_G/g_G^0$ and the gap $\Delta_e$. Parameters are: $V_Q=4 N_s U$, $N_b=N_s=6$, $g_G^0=20$.
}}}
\label{Cov}
\end{figure}

\begin{figure}[t]
\includegraphics[width=0.43\textwidth]{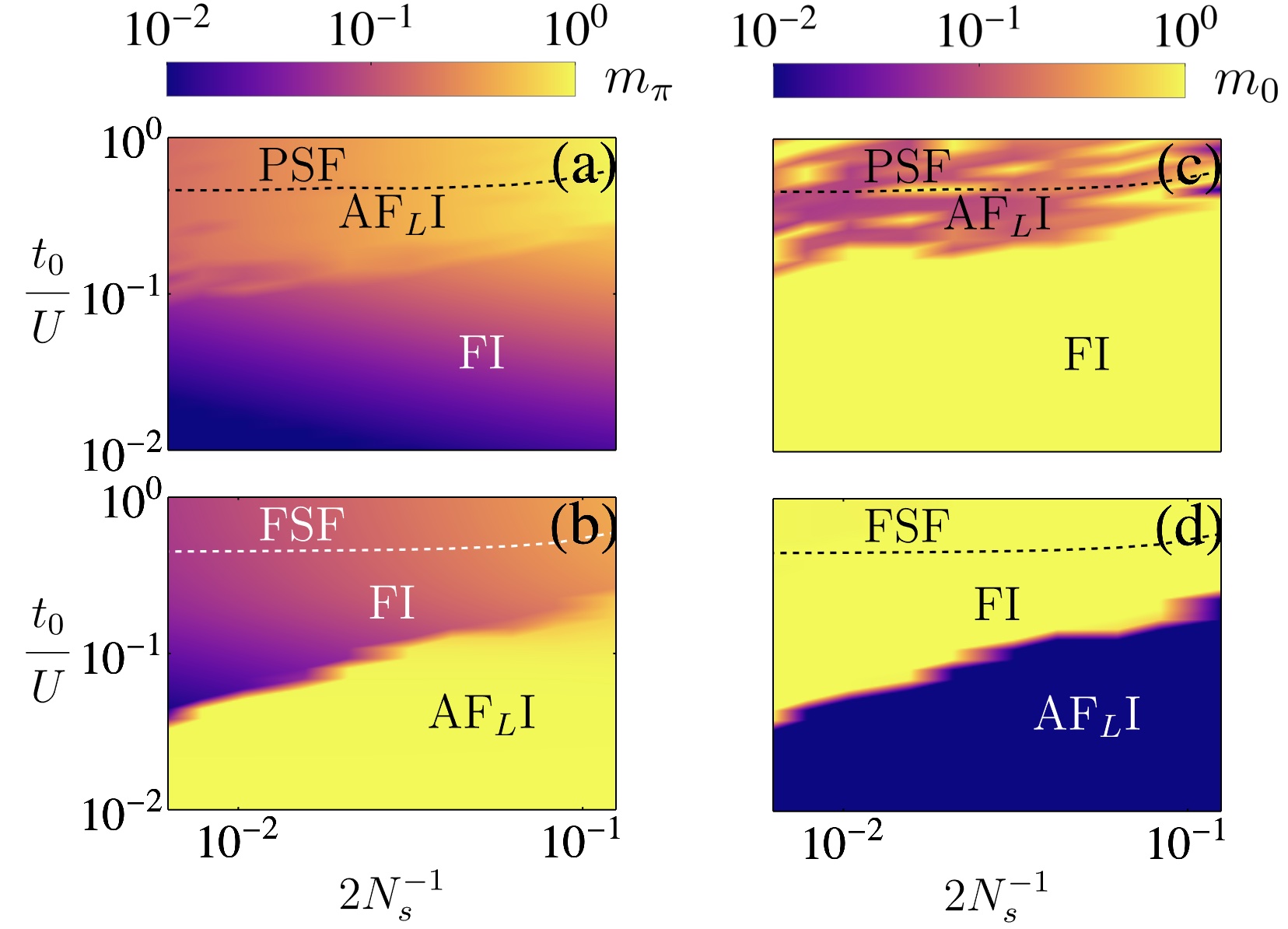} 
\captionsetup{width=0.43\textwidth}
\caption{
{\footnotesize{ 
{\it Staggered magnetization and magnetization}.
Panels (a,b) $m_\pi$ and (c,d) $m_0$ from DMRG simulations in 1D with $\phi_-$.  Estimated SF-MI boundary, dashed lines. 
Parameters are: (a,c) $V_Q=V_C=0.035U$; (b,d) $V_Q=V_C=-0.03U$ with maximum number of atoms per site $n_{max}=4$ and commensurate filling $N_b=N_s$.
}}}
\label{Mag}
\end{figure}

\textit{Numerical methods.}  
Computational simulations are performed with the effective Hamiltonian, equation~(3) in the main text. In our exact diagonalization (ED) simulations\cite{ed}, we use an optimized Lanczos scheme with sparse representation to find the ground states and the observables implemented with the ``Armadillo C++ library''~\cite{armadillo}. As the Hilbert space grows geometrically, we are limited to small sizes in ED. We consider lattices sizes of 6 ($2\times3$), 8 ($2\times4$) and 9 ($3\times3$) sites with three (effective 18, 24 and  27 sites) and also effectively two  (effective 12, 16, and 18 sites) spin components with periodic boundary conditions. The Hilbert space sizes range from $\sim 10^5$ to $\sim 10^7$ basis states. We explore different maximum number of atoms per site, and obtain similar qualitative results for $n_{max}=2$ to $n_{max}=N_s$ in ED.  The results for simulations do not change qualitatively as the number of sites change.  We perform additional simulations using density matrix renormalization group (DMRG) to verify that the gap and other observables are consistent with our ED computations. We perform 1D DMRG simulations using the ``itensor library''~\cite{itensor}. We have a 1D chain with $N_s$ from 4 to 98 with 2 spin components in a bosonic Hilbert space and open boundary conditions. We explore maximum number of atoms per site with fillings $n_{max}=2$ to $n_{max}=4$ and we find qualitative agreement between simulations. We also consider different initial conditions starting from either AF, F or random amplitude states with fixed commensurate fillings $\rho=N/N_s=1$ and we find the same results with different speeds of convergence to the ground state and first few exited states.  From finite size scaling in 1D DMRG, we find that the boundary between AF$\leftrightarrow$F transitions will have critical tunneling as $t_c\sim (a_0+a_1{N_s}^{-3/4})U$, with $a_0\sim 10^{-1}$ and $a_1\sim2$. 
The linear and quadratic magnetic shifts used are $\epsilon_\sigma\sim10^{-8}U$ for ED and $\epsilon_\sigma\sim10^{-4}U$ for DMRG. ED results do not change significantly with  $\epsilon_\sigma=10^{-8}U-10^{-3}U$ for small lattices. For larger values of $\epsilon_\sigma$ the boundaries of the AF$\leftrightarrow$F QPT shift but the transitions remain.

\textit{Results from ED for $V_C=0$, $V_Q\neq0$.}
AF states with global or local character present the same behaviour in the quantum covariance $|C_{\uparrow,\downarrow}|$. We show the results of $|C_{\uparrow,\downarrow}|$, the gap, the total number fluctuations and the degeneracy with ED  with 6 ($2\times3$) lattice sites, 2 spin components with periodic boundary conditions and $\phi_+$ for $V_Q>0$  in Fig.\ref{Cov}. The gap is calculated with respect to the ground state degenerate manifold, for $t_0/U=0$ the degeneracy of the ground state is $g_G^0$. For $N_s=N_b=6$, $g_G^0=20$. The total number fluctuations are scaled with respect to the $U=0$ limit, $\lim_{U\to0}\Delta(\hat{n}_i)^2=1-1/N_s$. For $t_0/U\gtrsim 1$, $\Delta_e\sim t_0/U$ when $V_Q\geq4U/N_s$ deep in the SF.

\textit{Magnetizations from DMRG 1D.}
The staggered magnetization $m_\pi$ and the magnetization $m_0$ for 1D DMRG simulations  in the  QPT AF$\leftrightarrow$F competition scenario for $\phi_-$ as a function of the number of sites is shown in Fig.\ref{Mag}. The behaviour of the magnetizations correlates with the behaviour of the gap shown in the main text.
 

\begin{thebibliography}{99}
\bibitem{Ferro}Note that the effective single component ferromagnetic system has a degenerate manifold with $g_F\sim \sqrt{\frac{3}{2}}\mathrm{exp}(3\log\left[\frac{4}{3}\right]N_s)$ states and an excitation gap deep in the MI: $\Delta_e=U+V_C$, a renormalized single component Bose-Hubbard model.
\bibitem{RDM}The reduced density matrix is $ 
 \rho_\sigma=\mathrm{Tr}_{\neq\sigma}[\rho]=\mathrm{Tr}_{-\sigma}[\mathrm{Tr}_0[\rho]]$, $\sigma\in\{\uparrow,\downarrow\}$. The    
partial trace is $\mathrm{Tr}_\sigma[\cdot]$, over all the lattice sites.
\end{thebibliography}
  \end{document}